\documentclass[prb,footinbib]{revtex4}
\usepackage{amssymb}

\input{tcilatex}

\begin{document}

\title{Reply to the comment by E. Men\'{e}ndez-Proupin on ``Multi-phonon
Raman scattering in semiconductor nanocrystals: Importance of non-adiabatic
transitions''}
\author{E. P. Pokatilov}
\affiliation{Department of Theoretical Physics, State University of Moldova, str. A.
Mateevici 60, MD-2009 Kishinev, Republic of Moldova}
\author{S. N. Klimin$^{\ast }$, V. M. Fomin$^{\ast }$ and J. T. Devreese$%
^{\ast \ast }$}
\affiliation{Theoretische Fysica van de Vaste Stof, Universiteit Antwerpen (U.I.A.),
B-2610 Antwerpen, Belgium}
\author{F. W. Wise}
\affiliation{Department of Applied Physics, Cornell University, Ithaca, New York 14853,
USA}
\date{May 23, 2002}

\begin{abstract}
In our recent paper [Phys. Rev. B \textbf{65}, 075316 (2002)], the selection
rules for the Raman scattering in spherical nanocrystals are analyzed. Here
we show, that a physically expedient choice of the coordinate system,
related to the polarization vector of the incoming or scattered light,
simplifies and clarifies the analysis.
\end{abstract}

\maketitle

In the paper \cite{Paper}, multiphonon Raman scattering in semiconductor
nanocrystals is investigated. In that work, we have considered two groups of
selection rules: (i) for active exciton states and (ii) for participating
phonons. Within the dipole approximation, the physically expedient choice
implies that the axis $Oz$ is parallel to one of the polarization vectors,
either $\mathbf{e}^{I}$ of the linearly polarized incoming light or $\mathbf{%
e}^{S}$ of the linearly polarized scattered light. The selection rules on p.
5 of the paper \cite{Paper} correspond just to this choice of the axis $Oz$.
We keep one and the same coordinate system when describing both exciton and
phonon states throughout the paper \cite{Paper}.

The selection rules for participating phonons can be expressed in terms of
the vector addition of angular moments. Let us denote $L_{I\left( S\right)
}\equiv 1$ the angular momentum of the incoming (scattered) light, and $l$
the angular momentum of a phonon. The corresponding $z$-projections (for an
arbitrary coordinate system) are $M_{I\left( S\right) }$ and $m,$
respectively. The rules for the angular-momentum addition are%
\begin{equation}
\left\{ 
\begin{array}{c}
\left| l-1\right| \leqslant 1, \\ 
M_{I}=M_{S}+m.%
\end{array}%
\right.  \label{0}
\end{equation}%
Herefrom, we obtain the selection rules for the phonon angular momentum and
for its $z$-projection%
\begin{eqnarray}
l &\leqslant &2,  \label{1} \\
m &=&M_{I}-M_{S}.  \label{2}
\end{eqnarray}%
Eq. (\ref{1}) means, in particular, that only $s$-, $p$- and $d$- phonons
may participate in the one-phonon Raman scattering. The selection rule (\ref%
{2}) is just the conservation law for the $z$-projection of the angular
momentum.

Let us consider the case when $Oz\parallel \mathbf{e}^{I},$ so that $%
M_{I}=0. $ For the Raman scattering in the parallel polarizations $\left( 
\mathbf{e}^{S}\parallel Oz\right) ,$ $M_{S}=0,$ and, according to Eq. (\ref%
{2}), only the scattering with the participation of phonons with $m=0$ is
allowed. Otherwise, when one considers the same Raman scattering in the
parallel polarizations choosing $Ox\parallel \mathbf{e}^{I}\parallel \mathbf{%
e}^{S},$ the projections of the angular moments for both incoming and
scattered photons on the axis $Oz$ do not take definite values, because
photon states in this case are superpositions of the states with $M_{I\left(
S\right) }=1$ and $M_{I\left( S\right) }=-1.$ For this coordinate system,
following Eq. (\ref{2}), phonons with $m=0,\pm 2$ can participate in the
one-phonon Raman scattering. This result is in agreement with the statement
of Ref. \cite{Comment}, that in parallel polarizations, phonons with $m\neq
0 $ can participate in the Raman scattering when the axis $Oz$ does not
coincide with the light polarization vectors. However, in this case, the
choice of the axis $Oz$ is not physically expedient.

For the scattering in the crossed polarizations, let us consider one of the
aforesaid physically expedient coordinate systems with $\left( Oz\parallel 
\mathbf{e}^{I},Ox\parallel \mathbf{e}^{S}\right) $. In this system, $%
M_{I}=0, $ while the scattered light is a superposition of states with $%
M_{S}=\pm 1.$ Consequently, the selection rule (\ref{2}) takes the form $%
m=\pm 1.$ It is worth noting, that for any coordinate system with $%
Oz\parallel \mathbf{e}^{I} $ or $Oz\parallel \mathbf{e}^{S}$, we obtain the
same selection rule $m=\pm 1 $ for the crossed polarizations.

For the other coordinate system $\left( Ox\parallel \mathbf{e}%
^{I},Oy\parallel \mathbf{e}^{S}\right) $, described in the comment \cite%
{Comment}, the photon angular momentum projections are: $M_{I}=\pm 1,$ $%
M_{S}=\pm 1$. As a result, $m$ can take the values $m=0,\pm 2.$ It is worth
mentioning, that when one chooses $Ox\parallel \mathbf{e}^{I}\parallel 
\mathbf{e}^{S}$ for the parallel polarizations and $Ox\parallel \mathbf{e}%
^{I},Oy\parallel \mathbf{e}^{S}$ for the crossed polarizations, the
selection rule (\ref{2}) leads to the same allowed values of $m$ for the
parallel and crossed polarizations, what makes a distinction between these
two scattering configurations less clear.

In summary, in different coordinate systems the selection rule (\ref{2}) for
the participating phonons (as well as the selection rules for the active
exciton states) can formally distinguish just because a rotation of the axis 
$Oz$ can lead to superpositions of states with different values of $m$ in a
new coordinate system. As shown above, the choice of the coordinate system
in the paper \cite{Paper} is physically opportune. When a coordinate system
is chosen in another way, the content of scattering can be masked, though it
describes the same physical processes.


\begin{thebibliography}{{{{{{{{{{{{{{{{{{{*}}}}}}}}}}}}}}}}}}*}
\bibitem[*]{A1} Permanent address: Department of Theoretical Physics, State
University of Moldova, str. A. Mateevici 60, MD-2009 Kishinev, Republic of
Moldova

\bibitem[{{{{{{{{{{{{{{{{{{*}}}}}}}}}}}}}}}}}}*]{A2} Also at: Universiteit
Antwerpen (RUCA), Groenenborgerlaan 171, B-2020 Antwerpen, Belgium and
Technische Universiteit Eindhoven, P. B. 513, 5600 MB Eindhoven, The
Netherlands

\bibitem{Paper} E. P. Pokatilov, S. N. Klimin, V. M. Fomin, J. T. Devreese,
and F. W. Wise, Phys. Rev. B \textbf{65}, 075316 (2002).

\bibitem{Comment} E. Menendez-Proupin, cond-mat/0204502 (unpublished).
\end{thebibliography}
\end{document}